\documentclass[iop,apj,numberedappendix,appendixfloats]{emulateapj}

\graphicspath{{./}{figures/}}
\DeclareGraphicsExtensions{.eps,.ps}
\usepackage{ulem}
\usepackage{xcolor}
\usepackage{natbib} 
\usepackage{multirow}

\shorttitle{Wide bandwidth observations of pulsars C, D and J in 47 Tucanae}
\shortauthors{Lei Zhang et al.}

\begin{document}
\title{Wide bandwidth observations of pulsars C, D and J in 47 Tucanae}

\email{leizhang996@nao.ac.cn}
\email{dili@nao.cas.cn}

\author{Lei Zhang$^{1,2,3*}$, George Hobbs$^{3,1}$, Richard N. Manchester$^{3}$, Di Li$^{1,2,4*}$, Pei Wang$^{1}$, Shi Dai$^{3,1}$, Jingbo Wang$^{5}$, Jane F. Kaczmarek$^{3}$,  Andrew D. Cameron$^{3,1}$, Lawrence Toomey$^{3}$, Weiwei Zhu$^{1}$, Qijun Zhi$^{6,7}$, Chenchen Miao$^{1,2}$, Mao Yuan$^{1,2}$, Songbo Zhang$^{2,3,8,9}$, Zhenzhao Tao$^{6,7,1}$  }

\affiliation{
$^1$ National Astronomical Observatories, Chinese Academy of Sciences, A20 Datun Road, Chaoyang District, Beijing 100101, China\\
$^2$ University of Chinese Academy of Sciences, Beijing 100049, China\\
$^3$ CSIRO Astronomy and Space Science, PO Box 76, Epping, NSW 1710, Australia\\
$^4$ FAST Collabration, CAS Key Laboratory of FAST, NAOC, Chinese Academy of Sciences, Beijing 100101, China\\
$^5$ Xinjiang Astronomical Observatory, 150, Science-1 Street, Urumqi, 830011 Xinjiang,  China\\
$^6$ Guizhou Provincial Key Laboratory of Radio Astronomy and Data Processing, Guizhou Normal University, Guiyang, China, 550001\\
$^7$ School of Physics and Electronic Science, Guizhou Normal University, Guiyang, China, 550001\\
$^8$ Purple Mountain Observatory, Chinese Academy of Sciences, Nanjing 210008, China\\
$^9$ International Centre for Radio Astronomy Research, University of Western Australia, Crawley, WA 6009, Australia\\
}

\begin{abstract}
We report the first wideband observations of pulsars C, D and J in the globular cluster 47\,Tucanae (NGC 104) using the Ultra-Wideband Low (UWL) receiver system recently installed on the Parkes 64\,m radio telescope. The wide frequency range of the UWL receiver (704--4032\,MHz), along with the well-calibrated system, allowed us to obtain flux density measurements and polarization pulse profiles. The mean pulse profiles have significant linear and circular polarization, allowing for determination of the Faraday rotation measure for each pulsar. Precise measurements of the dispersion measures show a significant deviation in the value for pulsar D compared to earlier results. Searches for new pulsars in the cluster are on-going and we have determined optimal bands for such searches using the Parkes UWL receiver system. 
\end{abstract}

\keywords{globular clusters: individual (47 Tucanae) -- pulsars: individual: J0024$-$7204C, J0024$-$7204D, J0024$-$7204J -- radio continuum: stars}

\section{Introduction}\label{sec:intr}
With 25 known radio pulsars, 47\,Tucanae (47\,Tuc or NGC 104) is the second most populous globular cluster (GC) in terms of number of known pulsars. All the pulsars were discovered using the Parkes 64\,m radio telescope at a wavelength of 20, 50 or 70\,cm \citep{Manchester90, Manchester91, Robinson95, Camilo00, Lorimer03, Knight07, Pan16}. All 25 are millisecond pulsars  with spin periods smaller than 8\,ms and 15 are in binary systems. \citet{McConnell00} made Australia Telescope Compact Array (ATCA) radio images of the cluster at 1.4 and 1.7\,GHz, they detected 11 point sources within 5\,arcmin of the cluster centre and the positions of three of them can be identified with 47\,Tuc C, D and J \citep{Freire01}\footnote{Also known as PSRs~J0024$-$7204C, J0024$-$7204D and J0024$-$7204J.}. Measured flux densities are extremely variable, mainly because of scintillation \citep{Rickett77}, but also because some of the binary pulsars show eclipses of the radio emission by circumstellar gas (e.g. 47\,Tuc J; see \citealp{Robinson95} for more details). \citet{Camilo00} reported that most pulsars in 47\,Tuc are visible only when the pulsar signals are amplified by scintillation. They estimated mean flux densities of the pulsars in 47\,Tuc at 20\,cm by assuming that scintillation effects are similar for all pulsars. To date, 14 pulsars in 47\,Tuc have known flux densities measured at 436\,MHz, 640\,MHz and/or 1400\,MHz from the Parkes telescope (\citealp{Robinson95,Camilo00}).  47\,Tuc C, D and J have flux densities measured at 1400 and 1700\,MHz from ATCA \citep{McConnell00}. The most recent timing models for 22 of the known pulsars in the cluster were presented by \citet{Ridolfi16} and \citet{Freire17} using data taken in the 20\,cm (1400\,MHz) observing band with the multibeam receiver on the Parkes radio telescope \citep{SSmith96}. There are no previous wideband or well-calibrated polarization observations of pulsars in 47\,Tuc. 

Here we present observations of 47\,Tuc using the Ultra-Wideband Low (UWL) receiver system recently installed at the Parkes radio telescope \citep{Hobbs19}. This receiver covers the band 704 to 4032\,MHz with excellent sensitivity and polarization properties. We report wideband flux density measurements and spectral properties (Section~\ref{sec:flux}), updated measurements of dispersion measures (DMs) (Section~\ref{sec:DM}) and polarization pulse profiles along with Faraday rotation measures (RMs) (Section~\ref{sec:pol}) for pulsars C, D and J in 47\,Tuc.

\section{Observations and Data Reduction }\label{sec:obs}


Pulsar observations of 47\,Tuc have often used a ``search" mode, where multiple frequency channels are rapidly sampled, typically with sample intervals of order 100~$\mu$s. This enables both searches for previously unknown pulsars in the cluster and the study of multiple pulsars with a given observation. However, such observations generally have restricted pulse phase, frequency and polarisation information and often use few-bit digitisation, reducing their sensitivity and dynamic range. In contrast, we can observe a specific pulsar in the cluster using the fold-mode capability. This leads to a better defined pulse profile that can be accurately calibrated in both polarisation and flux density. We have therefore obtained multiple, fold-mode observations of three of the brightest 47\,Tuc pulsars (C, D and J) that cover the entire band of the UWL receiver.  

 A summary of the observations is given in Table~\ref{tb:psrCDJ_obs}, which lists the pulsar name, the observing date, the observation length, the number of pulse phase bins and the corresponding file name. The data for all observations were coherently de-dispersed and channelised with  1\,MHz channels, integrated for 20\,s and then written to disk. Each observation was preceded by a short (1~minute) observation with a pulsed calibration signal injected into the low-noise amplifiers \citep[see][for more details]{Hobbs19}. Some early observations had 1024 bins per period; these were reduced to 512 bins as a first stage of the signal processing for consistency with the rest of the data sets. We removed 5\,MHz at each edge of each of the 26 sub-bands to avoid the effects of aliasing \citep[see][for more details]{Hobbs19} and then manually excised data affected by radio frequency interference (RFI) in frequency and time for each channel and sub-integration. 
 
 To transform the measured intensities to absolute flux densities, we used observations of the radio galaxy 3C~218 (Hydra A; see \citealp{Xie19}, for more details).  We measured the flux density for each pulsar by first forming noise-free standard templates using the \textsc{psrchive} program \textsc{paas} and then using \textsc{psrflux} to cross-correlate the observed profile with the standard template to obtain a scaling factor and then the average flux density. The uncertainty of the flux density is estimated as the root-mean-square (rms) noise of the profile baseline. 
 
 We calibrated the polarisation response of the UWL feed using multiple observations of the bright millisecond pulsar PSR~J0437$-$4715 that covered a wide range of parallactic angles \citep{van04}. To probe the polarimetric properties for each pulsar, we created a weighted sum of the 128-MHz sub-band pulse profiles. Determining the signal-to-noise (S/N) of a low-S/N profile is prone to biases.  Therefore, sub-band pulse profiles in which the ${\rm S/N}$ was less than $10$ were not included in the summation. The pulsars are weak in the higher bands and so our final pulse profiles covered a band between 704 and 2240\,MHz. We determined the RM of the pulsars over this 1536\,MHz bandwidth, using the \textsc{rmfit} package. The measured RMs were then used to refer all PAs to the overall band centre, 1472\,MHz, before summing to form the polarisation profiles. 
 
 In order to determine the DM for each pulsar, we generated times of arrival (ToAs) from all combined observations. These ToAs were obtained by splitting the profiles into 13 unequal sub-bands defined by four groups: seven sub-bands from 704 to 1728\,MHz, two sub-bands from  1728 to 2240\,MHz, two sub-bands from  2240 to 3264\,MHz and one sub-band from 3264 to 4032\,MHz. We used the same noise-free standard template that was used when measuring the flux densities. The \textsc{tempo2} \citep{Hobbs06} software package was subsequently used to fit for the DM. 

The observations from the Parkes radio telescope have been obtained under project codes P1006 and PX500. Conditional on data embargoes, these data are available on CSIRO's data archive\footnote{https://data.csiro.au/} \citep{Hobbs11} -- the raw P1006 and PX500 data have an 18-month embargo period. We have also produced a publicly-downloadable data collection containing our processed data files\footnote{https://doi.org/10.25919/5d8d7c2ee516a} \citep{Zhang19}. This data collection contains (1) the raw files of multiple, fold-mode observations, (2) the calibration files, (3) the calibrated and RFI-removed profiles and (4) the flux and polarisation calibrated files for these three pulsars.

\begin{table}[]
\centering
\caption{Observations of 47\,Tuc C, D and J.}\label{tb:psrCDJ_obs}
\begin{tabular}{p{28 pt}ccp{15 pt}l}
\hline
Pulsar                     & Date & Duration & \multicolumn{1}{c}{$N_{\rm bin}$} & \multicolumn{1}{c}{Filename}                      \\
                           & (MJD)    & (s)      &           &                               \\ \hline
\multirow{8}{*}{47Tuc C}   & 58583.13 & 3348     & 1024      & uwl\_190410\_031517\_b4.rf    \\
                           & 58586.93 & 7278     & 1024      & uwl\_190413\_222147\_b4.rf    \\
                           & 58602.77 & 10430    & 512       & uwl\_190429\_183836\_b4.rf    \\
                           & 58671.82 & 4114     & 512       & uwl\_190707\_194346\_b4\_0.rf \\
                           & 58671.86 & 4106     & 512       & uwl\_190707\_194346\_b4\_1.rf \\
                           & 58677.67 & 5213     & 512       & uwl\_190713\_161547\_b4\_0.rf \\
                           & 58677.73 & 5220     & 512       & uwl\_190713\_161547\_b4\_1.rf \\
                           & 58677.79 & 5232     & 512       & uwl\_190713\_161547\_b4\_2.rf \\ \hline
\multirow{10}{*}{47 Tuc D} & 58583.22 & 2478     & 1024      & uwl\_190410\_052207\_b4.rf    \\
                           & 58587.12 & 3734     & 512       & uwl\_190414\_025617\_b4.rf    \\
                           & 58602.89 & 5529     & 512       & uwl\_190429\_213437\_b4.rf    \\
                           & 58669.66 & 4040     & 512       & uwl\_190705\_160226\_b4.rf    \\
                           & 58669.78 & 4510     & 512       & uwl\_190705\_184456\_b4.rf    \\
                           & 58674.79 & 5384     & 512       & uwl\_190710\_190536\_b4\_0.rf \\
                           & 58674.85 & 5402     & 512       & uwl\_190710\_190536\_b4\_1.rf \\
                           & 58677.86 & 4014     & 512       & uwl\_190713\_203906\_b4\_0.rf \\
                           & 58677.90 & 4020     & 512       & uwl\_190713\_203906\_b4\_1.rf \\
                           & 58677.95 & 4016     & 512       & uwl\_190713\_203906\_b4\_2.rf \\ \hline
\multirow{8}{*}{47 Tuc J}  & 58663.76 & 4463     & 512       & uwl\_190629\_181617\_b4\_0.rf \\
                           & 58663.81 & 4480     & 512       & uwl\_190629\_181617\_b4\_1.rf \\
                           & 58663.86 & 4506     & 512       & uwl\_190629\_181617\_b4\_2.rf \\
                           & 58672.72 & 5613     & 512       & uwl\_190708\_172027\_b4\_0.rf \\
                           & 58672.78 & 5620     & 512       & uwl\_190708\_172027\_b4\_1.rf \\
                           & 58672.85 & 5602     & 512       & uwl\_190708\_172027\_b4\_2.rf \\
                           & 58678.67 & 5134     & 512       & uwl\_190714\_160846\_b4\_0.rf \\
                           & 58678.73 & 5136     & 512       & uwl\_190714\_160846\_b4\_1.rf \\ \hline
\end{tabular}
\end{table}

\section{Results and discussion}\label{sec:res}

\subsection{Flux Density Measurements and Spectral Properties}\label{sec:flux}


\begin{figure}[]
\centering
\includegraphics[width=0.95\linewidth]{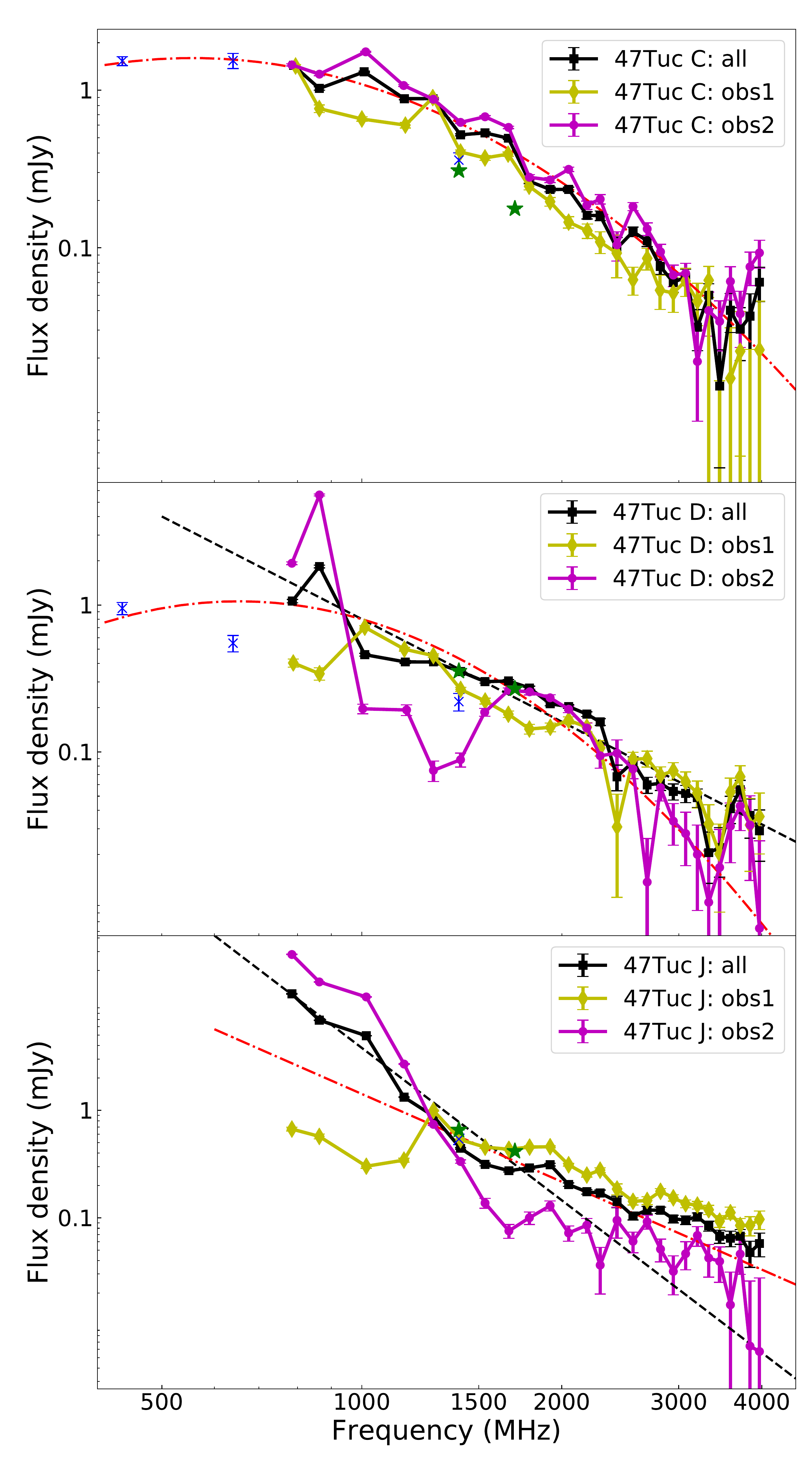}\\
\caption{Flux density as a function of frequency with 26 frequency sub-bands for 47\,Tuc C (upper panel), D (middle panel) and J (lower panel). The blue cross symbols are the previously measured flux densities at 436, 640 and 1400\,MHz for the pulsars from Parkes (\citealp{Robinson95, Camilo00}). The green star symbols are the previously measured flux densities at 1400 and 1700\,MHz for the pulsars from ATCA \citep{McConnell00}. The legend indicates the color/line scheme used to represent the average of all observation or specific observations.  The dashed and dot-dashed lines represent the spectral fitting results. See the text in Section~\ref{sec:flux} for more details.}
\label{fig.psrCDJ_flux}
\end{figure}

We present the flux density measurements for 47 Tuc C, D and J determined from 26 sub-bands (each with a 128\,MHz bandwidth) over the whole frequency band in  Table~\ref{tb:psrCDJ_inf} and shown in Figure~\ref{fig.psrCDJ_flux}. Each panel contains three solid lines.  For 47\,Tuc C, these correspond to flux densities from all combined observations (labelled ``all") the observations obtained in April (labelled ``obs1'') and the observations obtained in July (labelled ``obs2'').  For 47\,Tuc D and J the three lines correspond to the average of all observations (``all''), a sub-set of ``typical" observations (``obs1'') and the observation in which the pulsar was brightest (``obs2'').  For comparison, flux density estimates from previous work (\citealp{Robinson95, Camilo00, McConnell00}) are overlaid using blue cross and green star symbols.

These three pulsars have steep spectra. The spectra of 47\,Tuc C, can be best characterized as a log-parabolic spectrum (LPS). The form of the LPS is \citep{Jankowski18}:
\begin{equation}
\text{log}_{10}\text{S}_{\nu} = ax^{2}+bx+c,
\label{eq:PL_LP}
\end{equation}
where $x=\text{log}_{10}({\nu}/{\nu_{0}})$, $a$ is the curvature parameter, $b$ is the spectral index for $a=0$ and $c$ is a constant. The fitted parameters, $a$, $b$ and $c$ are $-$2.6(3), $-$4(2) and $-$2(5) where uncertainties in parentheses refer to the last quoted digit from $1\sigma$ level. The spectral fitting result is shown as the dot-dashed line in the top panel of Figure~\ref{fig.psrCDJ_flux}. The spectra of 47\,Tuc D can also be characterized as a LPS. The fitted parameters, $a$, $b$ and $c$ are $-$3.6(8), $-$4(7) and $-$1(2) (shown as the dot-dashed line in the middle panel of Figure~\ref{fig.psrCDJ_flux}).  We note that this LPS fit is strongly affected by the low-frequency flux density measurements at 436 and 640\,MHz.  If we only use the observations described in this paper, then the spectrum can be characterized as a simple power law \citep{Jankowski18}:
\begin{equation}
\text{S}_{\nu}=bx^{\alpha},
\label{eq:PL}
\end{equation} 
where $x = {\nu}/{\nu_{0}}$, $\alpha$ is the spectral index and $b$ is a constant. The fitted parameters, $\alpha$ and $b$ are $-$2.3(3) and 1(6). The spectral fitting result is shown as the dashed line in the Figure.

For 47\,Tuc J, we use a simple power law fitted to all sub-bands to measure a spectral index $\alpha$ = $-$4.7(3) with a constant $b$ = 1(7) (shown as the dashed line in the bottom panel of Figure~\ref{fig.psrCDJ_flux}). However, the flux density fluctuations between observations result in large uncertainties and affect the fitting for the spectral index. Therefore, for comparison, we also fitted the bottom 22 frequency sub-bands without the first four bright sub-bands and obtained a spectral index of $\alpha$ = $-$2.7(3) with a constant $b$ = 1(5). The spectral fitting result is shown as dot-dashed line in the Figure.

The results for 47\,Tuc D and J are not surprising as \citet{Jankowski18} reported that 79\% of pulsar spectra can be characterized as a simple power law. However, pulsars exhibiting a LPS spectrum, such as 47\,Tuc C, in this frequency band are much less common (accounting for about 10\% of the \citealt{Jankowski18} sample). 

\begin{figure}[]
\centering
\includegraphics[width=1.0\linewidth]{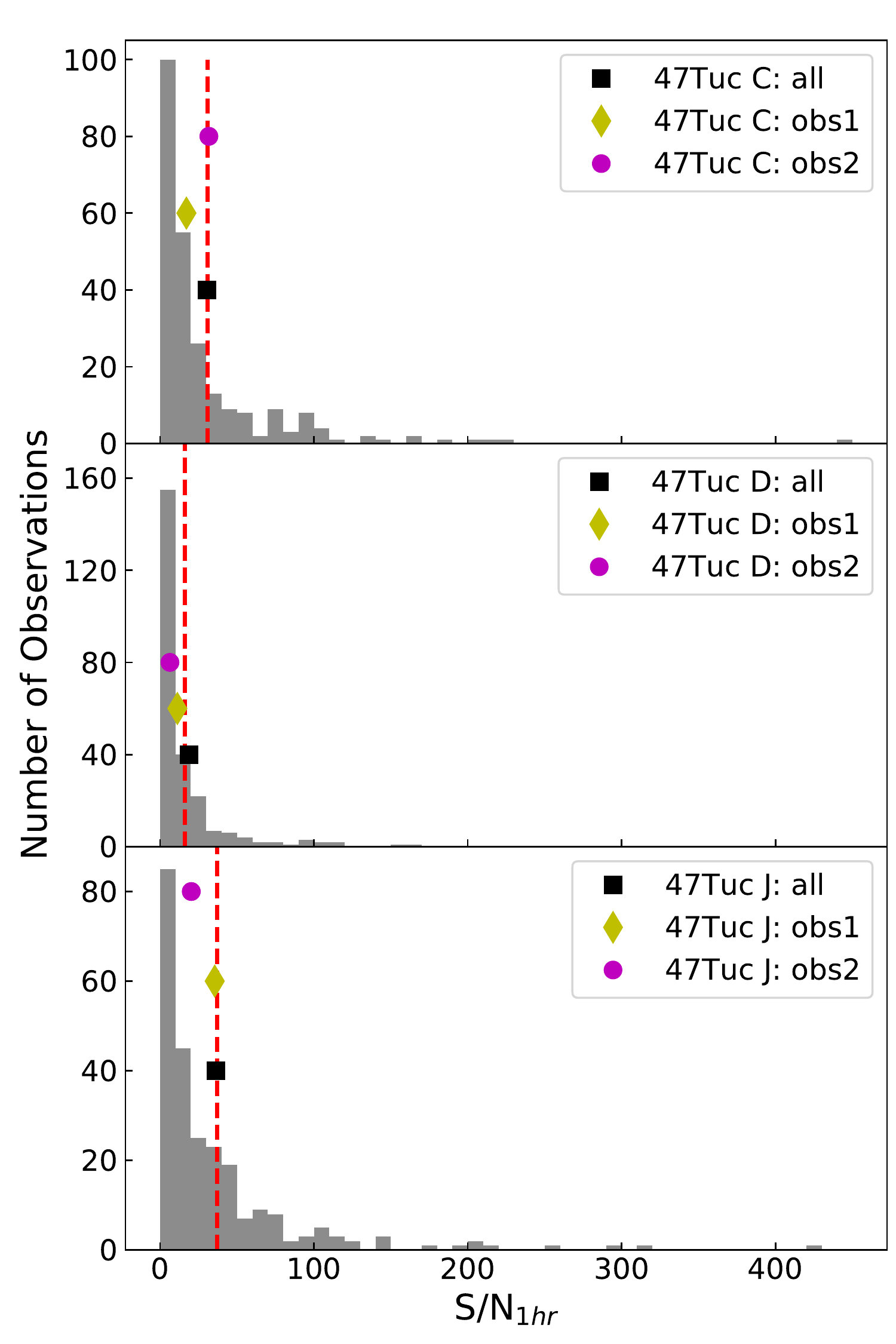}\\
\caption{Histograms of the signal-to-noise ratio (S/N) per hour for 47\,Tuc C (upper panel), D (middle panel) and J (lower panel). The distributions are the measured S/N at 1400\,MHz with 256\,MHz bandwidth for these three pulsars from previous Parkes surveys. Black square symbols show the S/N for all combined observations at 1400\,MHz with 256\,MHz bandwidth for the pulsars from this work. For 47\,Tuc C, the yellow diamond and purple circle symbols show combined observations on April and July. For 47\,Tuc D and J, the yellow diamond and purple circle symbols show combined observations from normal emission observations and bright observation. The red vertical dashed line marks the mean of S/N for the distribution of each pulsar.}
\label{fig.SN_dis}
\end{figure}

Scintillation is known to affect the measured flux densities of the pulsars in 47\,Tuc significantly (\citealp{Robinson95, Camilo00}) and our results show, in some sub-bands, flux density variations of more than an order of magnitude relative to the median value.  We can determine whether such variations are typical by comparing with the large number of observations of this cluster (spanning over 10 years) present in the CSIRO data archive \citep{Hobbs11}. We obtained 248 publically available observations for 47\,Tuc C, D and J obtained in the 20\,cm observing band with 256\,MHz bandwidth. The duration of the observation for each data file varied from 1 to 7 hours. We scaled the measured S/N of the folded profile of each pulsar to the value that would have been obtained after a 1\,hr observation.  The distribution functions of S/N variations for these three pulsars are shown in Figure~\ref{fig.SN_dis}; standard deviations are 47.3, 24.5 and 53.5 for 47\,Tuc C, D and J respectively. Our UWL results are consistent with these distributions. 

Searches for new pulsars in 47\,Tuc are on-going and the Parkes UWL receiver system is ideal for such searches. In particular, the signal processor allows search-mode data streams to be coherently de-dispersed at the known DM of the pulsars within the cluster. On average, the pulsars are brightest in the lowest frequency channels, but the flux densities also vary significantly in those bands.  At Parkes, the low frequency bands are also significantly affected by radio frequency interference (RFI). Using our observations we have identified clean bands and determined the S/N of the resulting profile in those bands. The S/N is lowest in the high frequency bands because the pulsars become weaker with increasing frequency.  As the bandwidth to search increases, the time taken to complete a search also increases and therefore as a compromise between computational feasibility, sensitivity and the ability to detect sources exhibiting a LPS spectrum, we suggest that 799--825\,MHz and 970--1260\,MHz are the optimal bands for independent pulsar searching with the UWL receiver at Parkes.

\begin{table}[]
\centering
\caption{Flux density measurements, dispersion measures， rotation measures and the width of pulse profiles at 50\% (W$_{50}$) and 10\% (W$_{10}$)  for 47\,Tuc C, D and J. Uncertainties in parentheses refer to the last quoted digit.}\label{tb:psrCDJ_inf}
\begin{tabular}{ccp{55 pt} llll}
\hline
                     & 47 Tuc C   & 47 Tuc D   & 47 Tuc J   \\ \hline
Flux Density (mJy)   &            &            &            \\
S$_{768}$            & 1.44(2)    & 1.06(3)    & 12.11(5)   \\
S$_{896}$            & 1.02(3)    & 1.8(4)     & 6.90(4)    \\
S$_{1024}$           & 1.31(1)    & 0.46(1)    & 4.93(1)    \\
S$_{1152}$           & 0.88(1)    & 0.41(1)    & 1.32(1)    \\
S$_{1280}$           & 0.88(1)    & 0.41(1)    & 0.88(1)    \\
S$_{1408}$           & 0.52(1)    & 0.35(1)    & 0.44(1)    \\
S$_{1536}$           & 0.53(1)    & 0.30(1)    & 0.31(1)    \\
S$_{1664}$           & 0.50(1)    & 0.30(1)    & 0.27(1)    \\
S$_{1792}$           & 0.26(1)    & 0.27(1)    & 0.29(1)    \\
S$_{1920}$           & 0.23(1)    & 0.21(1)    & 0.31(1)    \\
S$_{2048}$           & 0.23(1)    & 0.20(1)    & 0.20(1)    \\
S$_{2176}$           & 0.16(1)    & 0.18(1)    & 0.17(1)    \\
S$_{2304}$           & 0.16(1)    & 0.16(1)    & 0.17(1)    \\
S$_{2432}$           & 0.10(1)    & 0.07(1)    & 0.14(1)    \\
S$_{2560}$           & 0.12(1)    & 0.08(1)    & 0.10(1)    \\
S$_{2688}$           & 0.11(1)    & 0.06(1)    & 0.12(1)    \\
S$_{2816}$           & 0.08(1)    & 0.06(1)    & 0.12(1)    \\
S$_{2944}$           & 0.06(1)    & 0.05(1)    & 0.10(1)    \\
S$_{3072}$           & 0.07(1)    & 0.05(1)    & 0.10(1)    \\
S$_{3200}$           & 0.03(1)    & 0.05(1)    & 0.09(1)    \\
S$_{3328}$           & 0.05(1)    & 0.02(1)    & 0.08(1)    \\
S$_{3456}$           & 0.01(1)    & 0.02(1)    & 0.07(1)    \\
S$_{3584}$           & 0.04(1)    & 0.04(1)    & 0.06(1)    \\
S$_{3712}$           & 0.03(1)    & 0.06(1)    & 0.07(1)    \\
S$_{3840}$           & 0.04(1)    & 0.04(1)    & 0.05(1)    \\
S$_{3968}$           & 0.06(1)    & 0.02(1)    & 0.06(1)    \\\hline
DM (cm$^{-3}$ pc)    &            &            &            \\
Freire et al. (2017) & 24.600(4)  & 24.732(3)  & 24.588(3)  \\
This work            & 24.5955(8) & 24.7432(9) & 24.5932(1) \\\hline
RM (rad m$^{-2}$)$^{a}$    & 27(4)    & 28(4)    & 20(4)    \\\hline
W$_{50}$ (deg.)      &            &            &            \\
832 MHz              & 45.7       & 21.8       & 36.6       \\
1152 MHz             & 41.5       & 20.4       & 35.9       \\
1792 MHz             & 39.4       & 9.8        & 35.2       \\\hline
W$_{10}$ (deg.)      &            &            &            \\
832 MHz              & 75.9       & 49.2       & 63.3       \\
1152 MHz             & 73.8       & 48.5       & 56.3       \\
1792 MHz             & 71.7       & 46.4       & 54.8       \\ \hline
\end{tabular}
~~~~ $^{a}$ The RM uncertainties include the potential effect of ionospheric RM \citep{Sobey19}. 
\end{table}

\subsection{DM Measurements}\label{sec:DM}

We provide our measurements of the DM for 47\,Tuc C, D and J in Table~\ref{tb:psrCDJ_inf} along with previous determinations from \cite{Freire17}. Our DM determinations correspond to the central MJDs 58672, 58672 and 58673 for 47\,Tuc C, D and J respectively. Our DM measurements represent an improvement in precision over the previous measurements by about one order of magnitude. For 47\,Tuc C and J, the new values are not significantly different from those obtained by \citet{Freire17}  but for 47\,Tuc D the new DM deviates by more than $3\sigma$ from the previous measurement. Especially for 47\,Tuc D, the pulse profile evolves significantly with frequency (discussed in Section~\ref{sec:pol} below) and we investigated if this DM deviation could result from profile evolution over our wide band. Both the use of simulated synthetic profiles and separately deriving the DM from the lower two-thirds of the band and the upper two-thirds of the band showed that the effect of profile evolution on the derived DM was negligible, at least a factor of four smaller than the quoted DM uncertainties. The observed DM variations therefore appear real and are likely to have been caused by the moving line of sight traversing  electron-density irregularities in either the interstellar medium or the intracluster medium.  With continued wide-band observations, providing very precise DM estimates, spanning longer time spans we will be able to monitor such DM variations in detail.

\subsection{Polarization Pulse Profiles and RMs }\label{sec:pol}

\begin{figure*}[htp]
\centering
\includegraphics[width=18cm,angle=0]{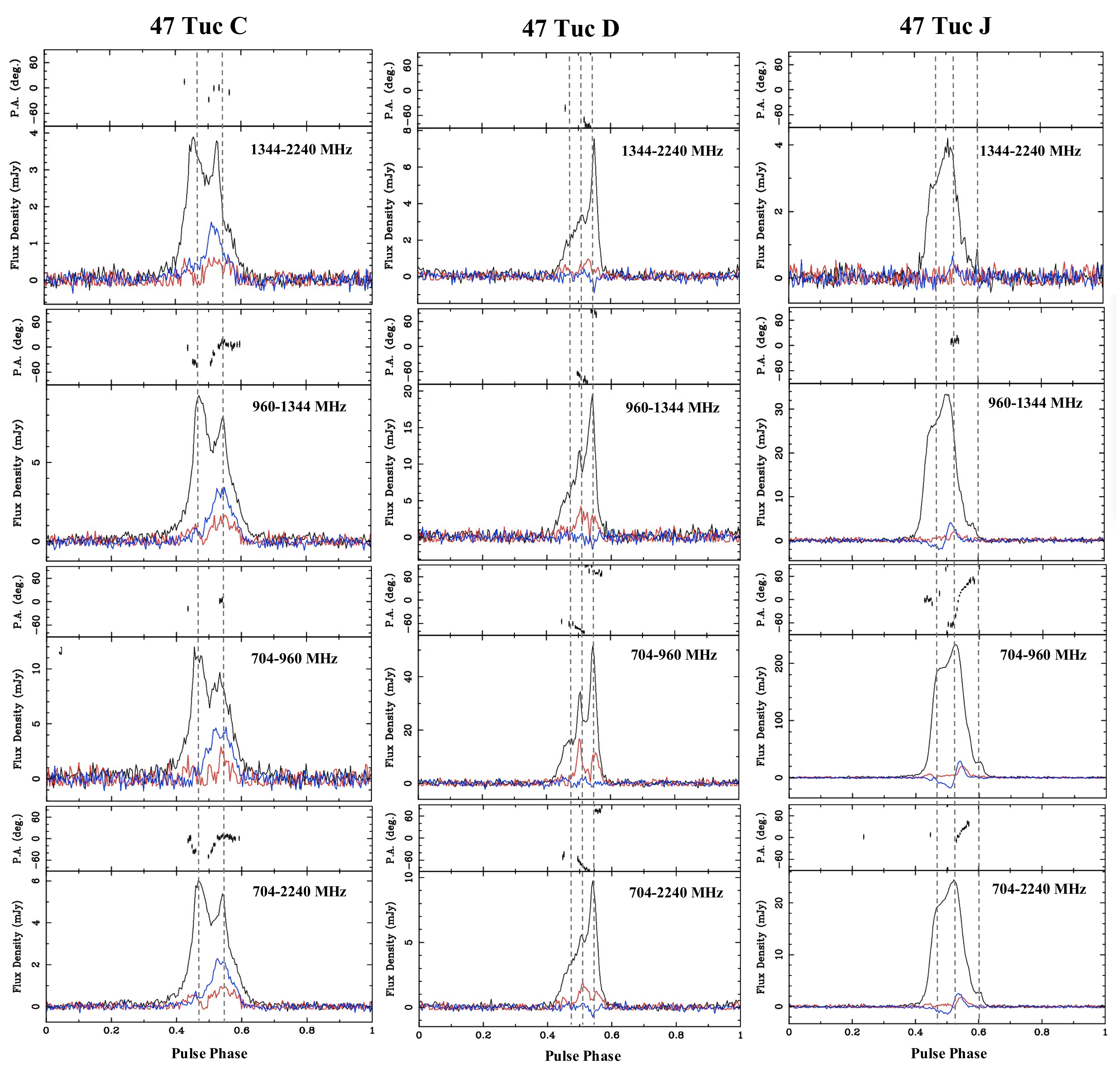}
\caption{Average polarization profiles for 47\,Tuc C, D and J in three different frequency bands between 704 and 2240\,MHz and the sum over this band. The black line is the mean flux profile, the red line is the linear polarization profile and the blue line is the circular polarization profile. Black dots in the top panel give the linear position angle (PA) referred to the overall band centre of the averaged profiles, 1472\,MHz.}
\label{fig.psrCDJ_pol}
\end{figure*}

In Figure~\ref{fig.psrCDJ_pol}, we present the mean polarization profiles of 47\,Tuc C, D and J in four different frequency bands (centred at 1792, 1152 and 832\,MHz for the top three panels). We also show the average profile for a much wider band centred at 1472\,MHz in the lowest panel.  We note that the S/N of the profiles in the UWL sub-bands above 2240\,MHz was too low to be included. The mean pulse profiles have significant linear (red line) and circular polarization (blue line). The RMs determined for each pulsar are listed in Table~\ref{tb:psrCDJ_inf}. 

For 47\,Tuc C, the observed pulse profiles show two components and the amplitudes of the two components become more equal as the observing frequency increases. The trailing component has stronger linear and circular polarization than the leading component. For 47\,Tuc D, the observed pulse profiles indicate three components, the leading and central components becoming relatively weaker compared to the trailing component as the observing frequency increases. Linear polarization is clearly seen in the central and trailing components.  For 47\,Tuc J, the pulse profiles show three components. Linear and circular polarization is only seen in the central component. 

Linear polarization and orthogonal-mode position angle (PA) jumps are commonly seen in pulsars (see e.g. \citealp{Ord04, Dai15}). In our sample we only see evidence for an orthogonal-mode PA jump between the first and second components of 47\,Tuc J (see the 704--960\,MHz band in Figure~\ref{fig.psrCDJ_pol}). The PAs of 47\,Tuc C (see the 930--1344\,MHz band) and J (see the 704--960\,MHz) vary significantly with pulse phase and they do not allow us to fit a `rotating vector model' (RVM; \citealp{Radhakrishnan69}). For 47 Tuc D (see the 704--960\,MHz), we can fit the RVM, however, we find that the magnetic inclination angle is unconstrained and therefore we cannot fully determine the emission geometry.
 
Over the wide observed band, we expect to see pulse shape evolution relating to intrinsic profile changes, emission arising from different positions in the magnetosphere, interstellar-medium and/or intracluster-medium effects. With coherently de-dispersed observations, we can obtain  pulse profiles that are not affected by inter-channel DM smearing. We measured the width of the profiles for 47\,Tuc C, D and J at 50\% and 10\% of the peak flux density in the three observing bands centred at 832, 1152 and 1792\,MHz (these results are listed in Table~\ref{tb:psrCDJ_inf}). As the observing frequency increases, the pulse widths decrease.  Such reduction in profile width is well-known in the general pulsar population and is usually attributed to radius-to-frequency mapping \citep{Cordes78}.

\section{Summary and conclusions}\label{sec:sum}

We have carried out the first wide bandwidth observations of pulsars C, D and J in 47\,Tuc. Flux density measurements and spectral properties (Section~\ref{sec:flux}), updated dispersion measurements (Section~\ref{sec:DM}) and polarization pulse profiles along with Faraday rotation measures (Section~\ref{sec:pol}) have been presented in this paper. We find small but significant changes in DM compared to previously published values, especially for 47\,Tuc D, which probably result from the motion of the line of sight across electron density irregularities. We will continue to observe the pulsars with the UWL receiver in order to monitor their DM and RM variations as well as building up higher S/N polarization profiles. The observations described here were independent observations for each pulsar, but clearly a significant increase in observational efficiency would occur if we could study all the pulsars simultaneously.  We can currently carry out such observations using search-mode observations, but the data volumes are prohibitive. We are therefore planning a new observing mode in which multiple pulsars (with similar DMs) can be folded online simultaneously.

\acknowledgments
\begin{acknowledgements}

This work is supported by the National Natural Science Foundation of China Grant No. 11988101, 11725313, 11690024, 11743002, 11873067, 11565010, 11603046, U1531246, U1731238, U1731218 and the Strategic Priority Research Program of the Chinese Academy of Sciences Grant No. XDB23000000.

QJZ is supported by the Science and Technology Fund of Guizhou Province (Grant Nos.(2015)4015, (2016)-4008, (2017)5726-37).
JBW is supported by the Youth Innovation Promotion Association of Chinese Academy of Sciences.

The Parkes radio telescope is part of the Australia Telescope National Facility which is funded by the Australian Government for operation as a National Facility managed by CSIRO. We thank the Parkes team for their great effort to install and commission the UWL receiver system. This paper includes archived data obtained through the CSIRO Data Access Portal\footnote{http://data.csiro.au}.
\end{acknowledgements}



\begin{thebibliography}{}
\bibitem[Camilo et al.(2000)]{Camilo00} Camilo, F., Lorimer, D.~R., Freire, P., et al.\ 2000, \apj, 535, 975

\bibitem[Cordes(1978)]{Cordes78} Cordes, J.~M.\ 1978, \apj, 222, 1006

\bibitem[Dai et al.(2015)]{Dai15} Dai, S., Hobbs, G., Manchester, R.~N., et al.\ 2015, \mnras, 449, 3223

\bibitem[Freire et al.(2001)]{Freire01} Freire, P.~C., Camilo, F., Lorimer, D.~R., et al.\ 2001, \mnras, 326, 901

\bibitem[Freire et al.(2017)]{Freire17} Freire, P.~C.~C., Ridolfi, A., Kramer, M., et al.\ 2017, \mnras, 471, 857

\bibitem[Hobbs et al.(2006)]{Hobbs06} Hobbs, G.~B., Edwards, R.~T., \& Manchester, R.~N.\ 2006, \mnras, 369, 655

\bibitem[Hobbs et al.(2011)]{Hobbs11} Hobbs, G., Miller, D., Manchester, R.~N., et al.\ 2011, PASA, 28, 202

\bibitem[Hobbs et al.(2019)]{Hobbs19} Hobbs, G., Manchester, R.~N., Dunning, A., et al.\ 2019, PASA, submitted

\bibitem[Jankowski et al.(2018)]{Jankowski18} Jankowski, F., van Straten, W., Keane, E. F., et al.\ 2018, \mnras, 473, 4436

\bibitem[Knight(2007)]{Knight07} Knight H. S., 2007, PhD thesis, Swinburne University of Technology

\bibitem[Lorimer et al.(2003)]{Lorimer03} Lorimer, D.~R., Camilo, F., Freire, P., et al.\ 2003, Radio Pulsars, 363

\bibitem[Manchester et al.(1990)]{Manchester90} Manchester, R.~N., Lyne, A.~G., D'Amico, N., et al.\ 1990, \nat, 345, 598

\bibitem[Manchester et al.(1991)]{Manchester91} Manchester, R.~N., Lyne, A.~G., Robinson, C., et al.\ 1991, \nat, 352, 219

\bibitem[McConnell \& Ables(2000)]{McConnell00} McConnell, D., \& Ables, J.~G.\ 2000, \mnras, 311, 841

\bibitem[Radhakrishnan \& Cooke(1969)]{Radhakrishnan69} Radhakrishnan, V., \& Cooke, D.~J.\ 1969, \aplett, 3, 225

\bibitem[Rickett(1977)]{Rickett77} Rickett, B.~J.\ 1977, \araa, 15, 479

\bibitem[Ridolfi et al.(2016)]{Ridolfi16} Ridolfi, A., Freire, P.~C.~C., Torne, P., et al.\ 2016, \mnras, 462, 2918

\bibitem[Robinson et al.(1995)]{Robinson95} Robinson, C., Lyne, A.~G., Manchester, R.~N., et al.\ 1995, \mnras, 274, 547

\bibitem[Ord et al.(2004)]{Ord04} Ord, S.~M., van Straten, W., Hotan, A.~W., et al.\ 2004, \mnras, 352, 804

\bibitem[Pan et al.(2016)]{Pan16} Pan, Z., Hobbs, G., Li, D., et al.\ 2016, \mnras, 459, L26

\bibitem[van Straten(2004)]{van04} van Straten, W.\ 2004, \apjs, 152, 129

\bibitem[Sobey et al.(2019)]{Sobey19} Sobey, C., Bilous, A.~V., Grie{\ss}meier, J.-M., et al.\ 2019, \mnras, 484, 3646

\bibitem[Staveley-Smith et al.(1996)]{SSmith96} Staveley-Smith, L., Wilson, W.~E., Bird, T.~S., et al.\ 1996, PASA, 13, 243

\bibitem[Xie et al.(2019)]{Xie19} Xie, Y.-W., Wang, J.-B., Hobbs, G., et al.\ 2019, Research in Astronomy and Astrophysics, 19, 103

\bibitem[Zhang et al.(2019)]{Zhang19} Zhang, L., et~al. 2019, Data files from Parkes v1. CSIRO. Data Collection, https://doi.org/10.25919/5d8d7c2ee516a\\
\end{thebibliography}
\end{document}